\documentclass[aps, preprint, showpacs, preprintnumbers, amsmath,
amssymb,prb]{revtex4}
\usepackage{graphicx}
\usepackage{dcolumn}
\usepackage{bm}
\usepackage{amsmath}
\usepackage{amsfonts}
\usepackage{amssymb}
\providecommand{\U}[1]{\protect\rule{.1in}{.1in}}

\begin{document}
\preprint{ }
\title[ ]{$^4$He Crystal Quality and Rotational Response in a
Transparent Torsional Oscillator}
\author{A. D. Fefferman, X. Rojas, A. Haziot, and S. Balibar}
\affiliation{Laboratoire de Physique Statistique de l'Ecole Normale Sup\'{e}rieure,
associ\'{e} au CNRS et aux Universit\'{e}s Denis Diderot et P.M Curie, 24 rue
Lhomond 75231 Paris Cedex 05, France}
\author{J. T. West and M. H. W. Chan}
\affiliation{Department of Physics, The Pennsylvania State University, University Park,
Pennsylvania 16802, USA}


\begin{abstract}
We have studied natural purity $^{4}$He single crystals and
polycrystals between 10 and 600 mK using a torsional oscillator with a
2 cm$^{3}$ rigid cell made of sapphire with a smooth geometry.  As the
temperature was lowered, we observed sample dependent but reproducible
resonant frequency shifts that could be attributed to a supersolid
fraction of order 0.1$\%$.  However, these shifts were observed with
single crystals, not with polycrystals.  Our results indicate that, in
our case, the rotational anomaly of solid helium is more likely due to
a change in stiffness than to supersolidity. This interpretation would
presumably require gliding of dislocations in more directions than previously thought.

\end{abstract}
\volumeyear{year}
\volumenumber{number}
\issuenumber{number}
\eid{identifier}
\received[Received text]{date}

\maketitle

\section{Introduction}
In 2004, Kim and Chan \cite{Kim2004,Kim2004a} discovered a rotational
anomaly in torsional oscillator (TO) measurements on solid $^{4}$He
and proposed that it was due to supersolidity.  In measurements of
the shear modulus of solid $^{4}$He, Day and Beamish \cite{Day2007}
observed a softening of the solid with the same dependence on
temperature and $^{3}$He concentration as that of the resonant
frequency shift in TO measurements.  The softening was attributed
to the evaporation of $^{3}$He impurities from dislocations as the
temperature $T$ increased above 100 to 200~mK. It was a surprise to
realize that, in its ``supersolid'' state where part of the mass is
supposed to flow through the rest of the solid, this solid was
actually stiffer than when no flow is observed.\cite{Balibar-Nature2010}
However, this apparent contradiction was tentatively explained by
assuming that mass superflow occurs in the core of dislocations only
if the latter are not free to move.\cite{Balibar-Nature2010}
According to this scenario, the rotational anomaly (i.e. the
supersolidity) and the elastic anomaly (the change in stiffness) could
be two consequences of a single phenomenon, namely the
binding/unbinding of $^3$He impurities to dislocations.

At the same time, one realized that, as $T$ was lowered, both an
inertia decrease due to the appearance of supersolidity and a
stiffening due to dislocation pinning should induce an increase of the
TO resonance frequency.\cite{Maris2011,Clark2008} Moreover,
J.D. Reppy observed a large frequency shift that was not consistent
with supersolidity and more likely due to a stiffness change.\cite{Reppy2010}  When a TO is made of several parts that are not
rigidly bound together, as is the case in Reppy's experiment, the TO
frequency shift may reach several percent of the ``mass loading'' -
the shift due to filling the TO with solid $^4$He.  We understand this
as a ``glue effect'', which means that solid $^4$He glues parts
together so that any change in the stiffness of solid $^4$He has a
large effect on the TO frequency. On the other hand, very
rigid TOs with a simpler geometry should not exhibit this glue effect
and the magnitude of the elastic effect has been much smaller.  This is the case with the
TO used by West et al.\cite{West2009} where the frequency shift was
about 0.015\% and the calculated contribution from the stiffness
change 60 times smaller.

As a consequence one should check in every experiment whether a
rotational anomaly in a TO measurement originates from a change in the
inertia or in the elastic coefficient (Ref. \onlinecite{Kim2012}).  For this reason we have
built a simple and rigid TO with a sapphire ``minibottle'' that could
be filled either with a single crystal or with a polycrystal (Fig.
\ref{schematic}).  If one assumes that superflow in a supersolid takes
place inside defects, the magnitude of the superfluid fraction should
increase with disorder and it should be {\it smaller} in single
crystals than in polycrystals.  This would be consistent with
measurements by Clark et al.\cite{Clark2007}  As for elastic
properties, Rojas \emph{et al}.\cite{Rojas2010} showed that single crystals
have a {\it larger} stiffness anomaly than polycrystals.\cite{Day2007}  The softening observed in the crystals
is consistent with a 50 to 86\% decrease in the elastic coefficient $c_{44}$ whereas that
in the polycrystals is due to a 7 to 15\% decrease in the effective shear modulus $\mu$.
Based on these numbers, one would expect a larger frequency shift for a single
crystal than for a polycrystal if the rotation anomaly is controlled by elasticity.\cite{Maris2011}  Our comparison of
single crystals to polycrystals in a sapphire TO could thus
discriminate between the two possible origins of rotational anomalies.
As we shall see below, we observed a larger effect with single
crystals, a result that is not consistent with the supersolid scenario.

We first verified that growth at constant volume from the normal
liquid using the ``blocked capillary method'' produces polycrystals
whose grain boundaries could be seen during melting, while growth from
the superfluid on the melting line produces single crystals without
grain boundaries and with easily visible
facets.\cite{Sasaki2008,Balibar2005,Pantalei2010}  As we shall see
below, we found that in our TO the rotation anomaly for a polycrystal
is below the resolution of our measurement, if it exists.  On the
opposite, we found definite frequency shifts for single crystals.  The
magnitudes of these frequency shifts were reproducible when measured
with the same crystal, but they varied from one single crystal to
another, as expected from recent calculations, if the crystal
orientation with respect to the rotation axis changed.\cite{Maris2011}

\section{Experiment}
\begin{figure}[ptb]
\centering
\graphicspath{{./figures/}}
\includegraphics[width=3.4in]{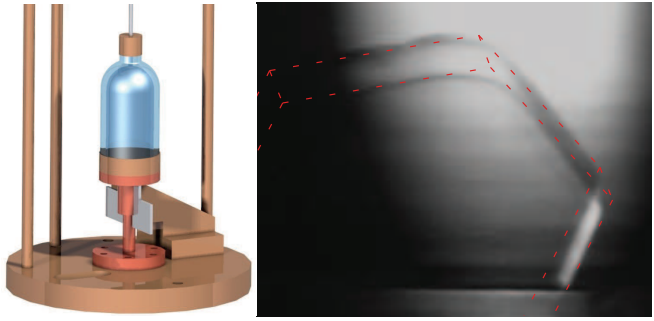}
\caption{(color) The TO (left) is filled from the top and is bolted to a
vibration isolation stage that is suspended from the dilution refrigerator
mixing chamber. For clarity, the second electrode structure (placed
diagonally from the displayed electrode structure) and the fourth suspension
rod are omitted from this drawing. Right panel: Photograph of an hcp single
crystal grown at 20 mK and 25.3 bar that partially fills the TO cell. The
image is dark at the edges due to refraction. The dashed red lines
illustrate the hexagonal shape of the crystal.}
\label{schematic}
\end{figure}

Our transparent sapphire minibottle sits atop a coin silver torsion
rod (Fig.~\ref{schematic}).  The sample volume in our cell has a
cylindrical geometry with a hemispherical part at the top to which the
fill line was glued with Stycast 1266.  This cell shape has no sharp
corners.  It was chosen to obtain a rigid cell that could be filled by
solid helium without leaving liquid regions behind during
crystallization when the liquid-solid interface goes up from the
bottom.  It also minimized the risk of facet blocking as the single
crystals reached the neck at the top of the minibottle.  The
cylindrical section had an inner diameter of 11.1 mm and a height of
15.2 mm.  The growth of polycrystals at constant volume started at
2.55 K and finished at 1.97 K, implying that the final pressure in the
solid was 38 bar.  This TO resonates near 910~Hz.  The ``mass
loading'' was 1.8~Hz. All measurements shown here were done with the fill
line of the 1K stage closed in order to minimize mechanical
vibrations.

We have grown many crystals in this cell.  For clarity we present the
last three ones for which the best stability and reproducibility was
obtained after optimization of our methods.  The sample labeled
``single crystal \#1'' was first grown at 10~mK with a growth speed
$\approx$10 $\mu$m/s.  Before any further treatment, the TO results in
this crystal were not reproducible as a function of time.  Random
jumps of order 1~mHz always occurred during the temperature scan (12
hours).  These jumps made it impossible to achieve the stable
resolution we needed (0.1 mHz).  Apparently, when a high quality
crystal is in equilibrium with liquid $^4$He at the orifice of the
fill line, some mass spontaneously rearranges somewhere, which makes
the TO frequency unstable.  We had to escape from the liquid-solid
equilibrium to obtain stable measurements.  This was achieved by
warming the crystal to 1.4 K, melting 10\% of the crystal and then
regrowing it with a growth speed $\approx$3 $\mu$m/s.  At 1.4~K, no
facet could block the growth near the orifice of the fill
line.\cite{Balibar2005}  Furthermore, if a small amount of liquid
still remained in the cell at 1.4 K, it would have solidified very
soon after cooling started.\cite{Sasaki2008}  Since cooling below
1.4~K along the melting line changes the cell pressure, some
anisotropic stresses may have produced some dislocations in this
single crystal until all the liquid space including the fill line had
solidified.  Single crystal \#2 was obtained in the same way but grown
from a different seed.  This crystal was warmed up to 1.45 K, at which
point 20\% of the crystal was melted and regrown.  In single crystals
grown at 1.4~K, Syshchenko and Beamish \cite{Syshchenko2008} measured
a dislocation density in the range 10$^{3}$ to 10$^{5}$ cm$^{-2}$.  We
estimate that this dislocation density is approximately equal to or possibly greater than the density in our single crystals.

\begin{figure}[ptb]
\centering
\graphicspath{{./figures/}}
\includegraphics[width=3.4in]{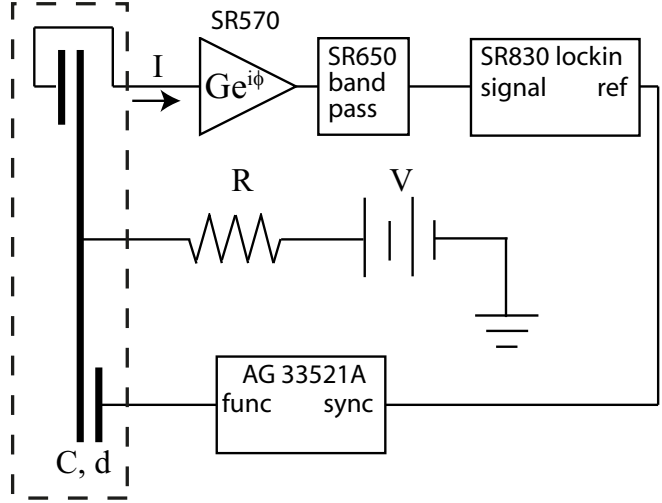}
\caption{The circuit used to measure the resonant frequency and Q
of the torsional oscillator. The components in the dashed box are at the
temperature of the mixing chamber.  The lockin amplifier and function generator are controlled via a GPIB interface.}
\label{cct}
\end{figure}

We primarily operated the TO at constant response amplitude and constant drive
frequency, extracting the resonant frequency, $f_0$, and quality factor, $Q$, from the
phase of the TO response and the drive level with a method similar to that of Ref. \onlinecite{Morley2002}. This method is
different from the more widely used phase locked loop where the response amplitude
and the drive frequency are not kept constant while scanning the
temperature.  We used the circuit shown in Fig. \ref{cct} for our measurement. The long
heavy black line in the dashed box represents the mobile electrodes of the TO and the short heavy
black lines represent the fixed electrodes.\ The spacing between the
electrodes, $d$, is 90 microns, yielding a capacitance $C=3$
pF.\ A bias voltage $V=$ 260 V is applied to the mobile electrodes via the
current limiting resistor $R=$ 2 M$\Omega $.\ A function generator is used
to apply an ac voltage, $\tilde{V}$, on the order of millivolts to the drive
electrode.\ The motion in response to the resulting torque modulates the
spacing and causes a current%
\begin{equation}
I=\frac{-i\omega CVl\theta }{d}  \label{current}
\end{equation}%
to flow into the current amplifier, since the voltage across the detection
capacitor remains constant.\ In Eq. \ref{current}, $\omega $ is the drive
angular frequency, $l=7.6$ mm is the radial position of the detection
electrode, and $\theta =\theta _{0}$e$^{i(\omega t+\alpha )}$ is the angular
position of the mobile electrodes, where $\alpha $ is the phase angle relative to $\tilde{V}%
.$\ The lockin amplifier then measures a voltage
\begin{equation}
V_{LI}=IGe^{i\phi },  \label{Vlockin}
\end{equation}%
where $G=10^{8}$ V/A is the gain of the current amplifier and $\phi $ is the
phase shift due to the measurement electronics.

The susceptibility $\chi \left( \omega \right) $ characterizes\ the
response, $\theta $, of a TO to an externally applied torque, $\tau $. Our
TO is well described by the form%
\begin{equation}
\chi ^{-1}=\tau /\theta =J\omega ^{2}-ib\omega -k,  \label{susceptibility}
\end{equation}%
where the coefficients $J,b$ and $k$ may be temperature dependent.\ We have
defined $\chi $ so that a positive static torque decreases $\theta $.\ The
resonant frequency $\omega _{0}$ and $Q$ of the TO are related to the root
of $\chi ^{-1}$ (Ref. \onlinecite{Graf2010}). We obtain%
\begin{equation}
\omega _{0}=\sqrt{\frac{k}{J}}
\end{equation}%
and%
\begin{equation}
Q^{-1}=\frac{b}{\omega _{0}J}.
\end{equation}%

\begin{figure}[ptb]
\centering
\graphicspath{{./figures/}}
\includegraphics[height=20cm]{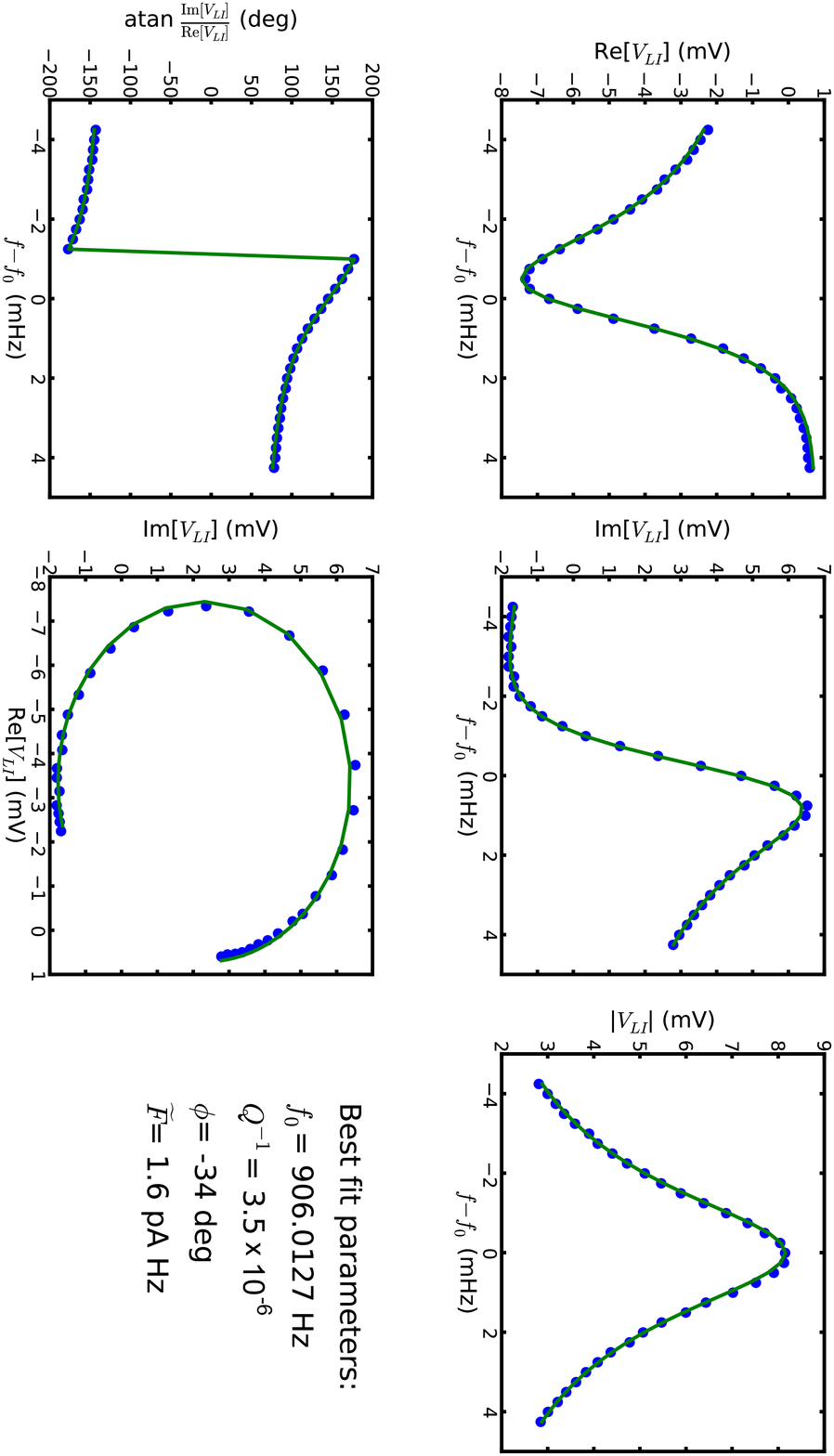}
\caption{(color online) Response of the torsional oscillator with the polycrystalline sample discussed in the text, measured using the circuit in Fig. \ref{cct}.}
\label{freq_sweeps}
\end{figure}

One way we determined $\omega _{0}$ and $Q$ of the TO was by measuring the
response of the TO at constant temperature and drive level while sweeping
the drive frequency through the resonance (a \textquotedblleft frequency
sweep\textquotedblright ). We fit the function%
\begin{equation}
V_{LI}=\frac{-i\tilde{F}\omega }{\left( \omega _{0}^{2}-\omega
^{2}+iQ^{-1}\omega \omega _{0}\right) }G\text{e}^{i\phi }  \label{response}
\end{equation}%
to the TO response as measured by the lockin amplifier. $\ $Eq. \ref%
{response} was derived from Eqs. \ref{current}, \ref{Vlockin}, and \ref%
{susceptibility}. The fitting parameters in Eq. \ref{response}\ are the
resonant frequency, $\omega _{0}$; the quality factor of the TO, $Q$; the
phase shift introduced by the measurement electronics, $\phi $; and $\tilde{F%
}$, which is proportional to $\tilde{V}$. An example of a frequency sweep is
given in Fig. \ref{freq_sweeps}. The best fit values of $\omega _{0}/2\pi $ and $Q^{-1}$
from the frequency sweeps are plotted as large filled circles in Figs. \ref{raw}-\ref{Qinv}.\ The color of the circle indicates that the frequency
sweep (at constant temperature) was carried out soon before or after the
temperature sweep with the same color.

The TO $f_{0}$ and $Q$ were measured at constant drive
frequency and constant response amplitude for the temperature sweeps.\
Substituting $\tau =\tau _{0}$e$^{i\left( \omega t+\pi \right) }$ and $%
\theta =\theta _{0}$e$^{i(\omega t+\alpha )}$ into Eq. \ref{susceptibility},
we obtain%
\begin{equation}
\tau _{0}\text{e}^{i\left( \pi -\alpha \right) }+k\theta _{0}+ib\omega
\theta _{0}-J\omega ^{2}\theta _{0}=0.  \label{eqofmotion}
\end{equation}%
Taking the real part of Eq. \ref{eqofmotion}, we obtain%
\begin{equation}
\omega _{0}^{2}=\omega ^{2}\left( 1+\frac{\tau _{0}\cos \alpha }{J\omega
^{2}\theta _{0}}\right) .  \label{resfreq}
\end{equation}%
We then make the approximation $k=J\omega ^{2}$ in Eq. \ref{resfreq}, since $%
\ (\omega -\omega _{0})/2\pi <20$ mHz for our measurements, yielding%
\begin{equation}
\omega _{0}^{2}=\omega ^{2}\left( 1+\frac{\dot{\tau}_{0}\cos \alpha }{k\dot{%
\theta}_{0}}\right) .  \label{resfreq2}
\end{equation}%
Furthermore, we note that $\tau _{0}/k=\theta _{R}/Q$, where $\theta
_{R}\equiv \theta _{0}$ at the resonant frequency. Since the relative
variation of $\omega _{0}$ over the entire temperature range of our
measurements is only $\approx 10^{-5}$, $k$ is nearly independent of
temperature. Thus, $\dot{\tau}_{0}/k\left\vert \tilde{V}\right\vert =\dot{%
\theta}_{R}/Q\left\vert \tilde{V}\right\vert =3.9\times 10^{-6}$ sec$^{-1}$V$%
^{-1}\equiv c$ is also nearly independent of temperature, and when this is
substituted into Eq. \ref{resfreq2}, we obtain%
\begin{equation}
\omega _{0}=\omega \sqrt{1+\frac{c\left\vert \tilde{V}\right\vert \cos
\alpha }{\dot{\theta}_{0}}}.  \label{freqfinal}
\end{equation}%
Taking the imaginary part of Eq. \ref{eqofmotion}, we obtain%
\begin{equation}
Q^{-1}=-\frac{\tau _{0}\sin \alpha }{J\omega _{0}\omega \theta _{0}}.
\end{equation}%
Using the same approximations that were used to obtain Eq. \ref{freqfinal},
we obtain%
\begin{equation}
Q^{-1}=-\frac{c\left\vert \tilde{V}\right\vert \sin \alpha }{\dot{\theta}_{0}%
}.  \label{Qfinal}
\end{equation}%
Holding $\dot{\theta}_{0}$ constant, we determined $\omega _{0}$ and $Q^{-1}$
by measuring $\alpha $ and $\left\vert \tilde{V}\right\vert $.

Before doing a temperature sweep, we set the drive frequency to minimize $%
\omega -\omega _{0}$ over the temperature range of the measurement. The
drive level was set so that the desired TO rim speed was approximately
obtained at the starting temperature of the temperature sweep.\ At this
point, we began to constantly acquire the value of $V_{LI}$ from the lockin
amplifier.\ The TO was allowed to equilibrate for 8 minutes at the initial
drive level.\ The values of $\omega _{0}$ and $Q^{-1}$ were then determined
by applying Eqs. \ref{freqfinal} and \ref{Qfinal} and averaging over the
next 4 minutes.\ The drive level was then adjusted, if necessary, to obtain
exactly the desired TO\ rim speed, and the TO\ was allowed to equilibrate
for 8 minutes again.\ The values of $\omega _{0}$ and $Q^{-1}$ at the
desired rim speed were then obtained by averaging for 10 minutes. The
temperature was then incremented or decremented, and the measurement routine
was repeated.

\section{Results}
\begin{figure}
\centering
\graphicspath{{./figures/}}
\includegraphics[width=3.4in]{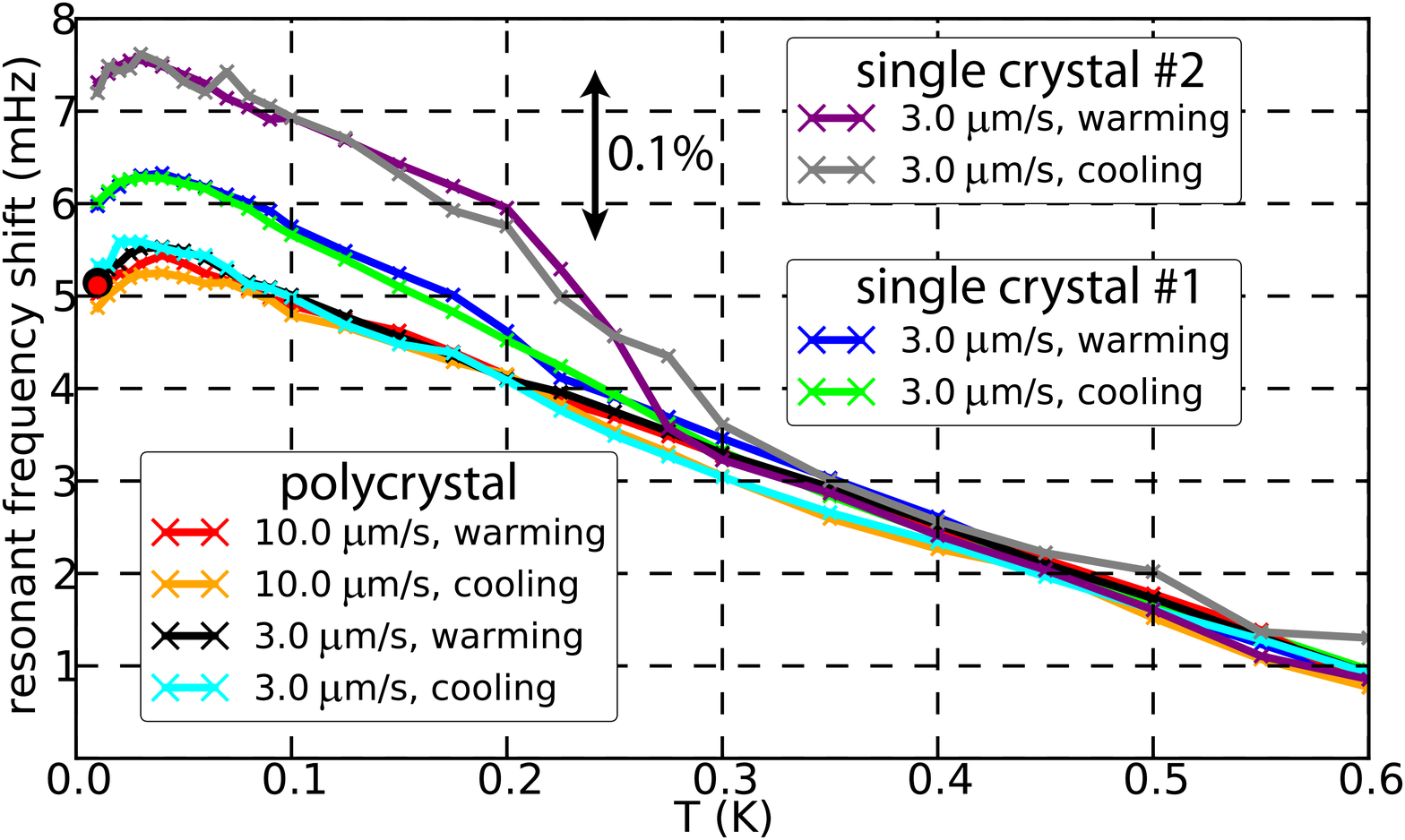}
\caption{(color) Temperature dependence of the resonant frequency shift
$\delta f$ for the polycrystal and two single crystals at the specified rim
speeds. For the polycrystal, $\delta f$ is linear above 80 mK, indicating
the absence of a rotation anomaly to within the resolution of our measurement
(0.2~mHz). The slightly amplitude dependent
curvature below 80 mK is due to the $T$ dependence of the TO fabrication
materials. For single crystals \#1 and \#2, $\delta f$ departs from the linear
$T$ dependence at intermediate temperatures. The red and black closed
circles at 10 mK that verify our constant drive frequency measurements were obtained by
fitting a Lorentzian to the TO resonant response. The scale bar relates the frequency shift to the TO mass loading (see text).}
\label{raw}
\end{figure}

In Fig.~\ref{raw} we present example measurements of the resonant
frequency shift $\delta f$ of the TO $versus$ temperature $T$ for the
three different samples.  This shift is the difference between the
actual TO frequency and a reference frequency.  This reference
frequency is adjusted for each sample so that the high temperature
parts superimpose.  The specified rim speeds are measured at the inner
surface of the cylindrical portion of the minibottle.  The frequency
shift with the polycrystalline sample is linear in $T$ above
$\approx$80 mK, as found when the cell is filled with superfluid except for a
different slope. Near 40~mK, the frequency shows a maximum which is
known to be due to the TO fabrication materials as found by other
groups.\cite{Morley2002} Thus, we see no evidence for a rotational anomaly
within the resolution of our experiment (0.2~mHz).  This means that,
if it exists, the superfluid fraction is less than 10$^{-4}$ in this
polycrystal.  It is also consistent with a change in $\mu$
of less than 15\%,\cite{Maris2011} a reasonable
value.\cite{Day2007}  This result is robust: for example, we
show in Fig.~\ref{raw} that the $T$ dependence of the frequency shift
is the same for rim speeds of 3 and 10 $\mu$m/sec for both warming and
cooling.  But for the single crystal samples, $\delta f$ departs from
the linear $T$ dependence in the intermediate temperature range.

\begin{figure}[ptb]
\centering
\graphicspath{{./figures/}}
\includegraphics[width=3.4in]
{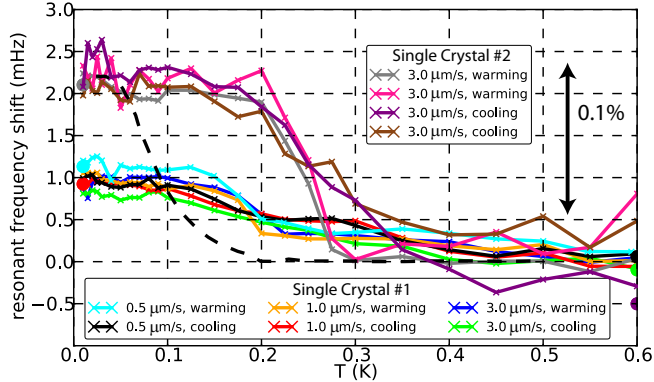}
\caption{(color) Temperature dependence of the TO resonant frequency shift
$\delta f$ for single crystals \#1 and \#2 relative to $\delta f_{\text{pc}}$ for the
polycrystal at the specified rim speeds. The $T$ dependence observed in most
other TO or elastic measurements on natural purity solid $^{4}$He, represented
by data from Ref. \onlinecite{Day2007} (vertically scaled), is shown with a dashed black
line for comparison. The six closed circles at 10 or 600 mK verify our measurements as
explained in the Fig. \ref{raw} caption. The scale bar relates the frequency shift to the TO mass loading (see text).}
\label{df_rel}
\end{figure}

The frequency shift for the two single crystal samples relative to the
frequency shift for the polycrystal $\delta f-\delta f_{\text{pc}}$\
is shown in Fig.  \ref{df_rel}.  We compute the frequency shift
relative to the polycrystal rather than a superfluid filled cell
because the latter has a linear $T$ dependence above 300 mK that is
different from the solid ones, as found by other
groups.\cite{Eunseong-thesis}  The slope is 9 mHz/K when the cell and
its fill line are filled with superfluid, compared with 8 mHz/K for
the solid samples.  Thus, computing $\delta f-\delta f_{\text{pc}}$
best exposes the relevant behavior.  As the temperature was decreased,
an increase in $\delta f-\delta f_{\text{pc}}$ was observed with a
magnitude of 1~mHz (2 mHz) for single crystal \#1 (\#2).  Our
observation of a frequency shift with single crystals which is sample
dependent but reproducible and definitely larger than with
polycrystals is qualitatively consistent with shifts
originating from a stiffness change, not an inertia change in our
samples.  In order to demonstrate reproducibility,
several temperature sweeps on both warming and cooling are
shown for each crystal.  There is no clear critical velocity over the
range of rim speeds studied here: for single crystal \#1, measurements
at 0.5, 1.0 and 3.0 $\mu$m/sec are the same within our
resolution.  A noticeable difference with other experiments is the
temperature where the shift occurs.  Half of it
was completed at 200 mK for single crystal \#1 and at
250 mK for single crystal \#2.  In contrast, other TO measurements or
direct elastic measurements of natural purity (300 ppb) solid helium
show that half of the frequency or stiffness shift is
completed at 90 mK, as shown by the black dashed line in
Fig.~\ref{df_rel} from Beamish's results.\cite{Day2007}

\begin{figure}[ptb]
\centering
\graphicspath{{./figures/}}
\includegraphics[width=3.4in]{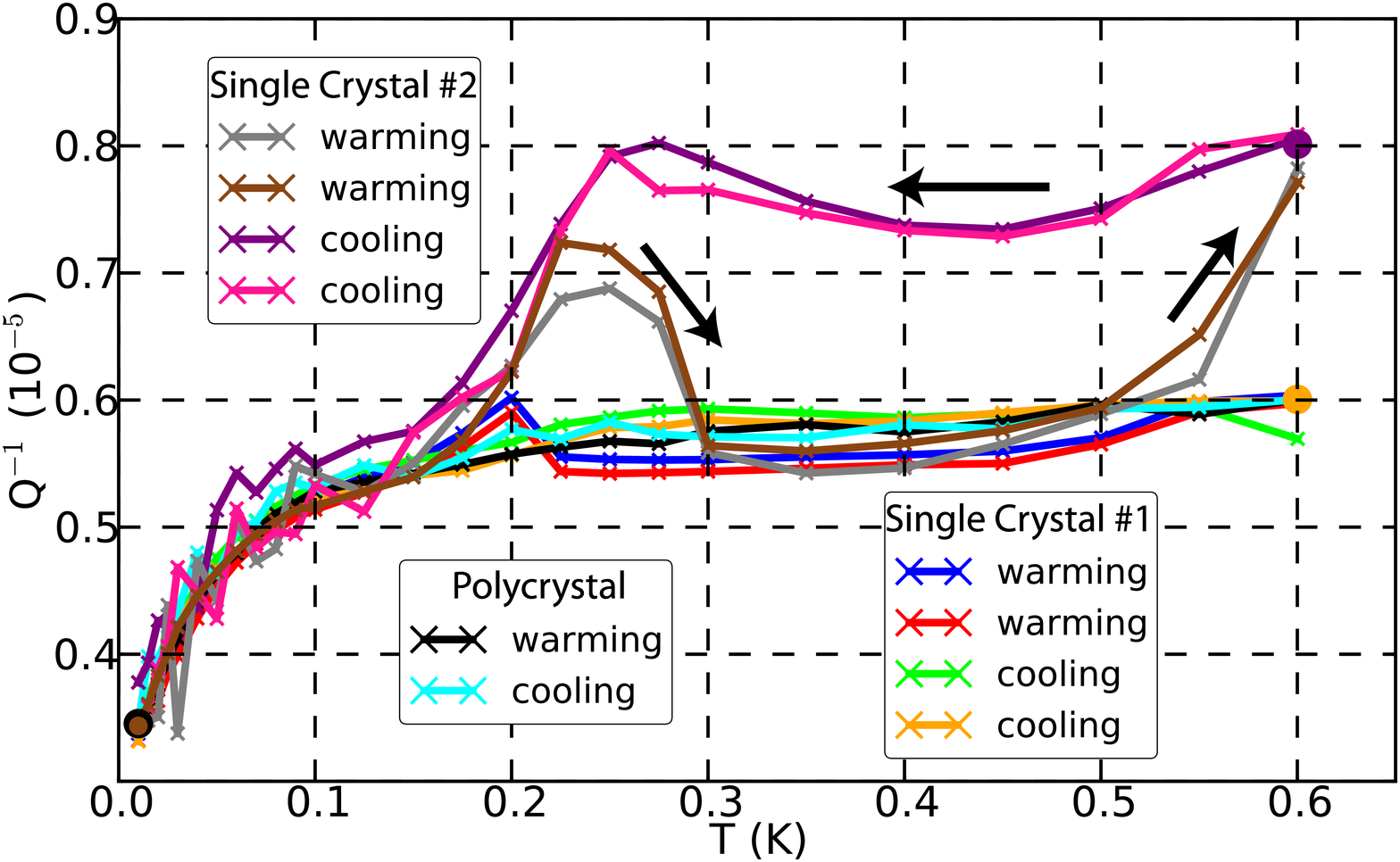}
\caption{(color) Temperature dependence of the TO dissipation for the
polycrystal and both single crystals measured at a rim speed of 3 $\mu$m/sec. The temperature of the prominent dissipation peak for single crystal \#2 on
warming coincides with the maximum of $|d(\delta f-\delta f_{\text{pc}})/dT|$ (Fig. \ref{df_rel}). An offset of $8\times10^{-7}$, probably due to a temporary increase in the TO background damping, was subtracted from one of the curves (purple). The four
closed circles at 10 or 600 mK verify our measurements as explained in the Fig. \ref{raw}
caption.}
\label{Qinv}
\end{figure}

The $T$ dependence of the TO dissipation $Q^{-1}$ is shown in Fig.
\ref{Qinv}. For the polycrystal, $Q^{-1}(T)$ has the standard $T$ dependence
generated by the TO fabrication materials: a temperature independent plateau
above 200 mK with a roll-off below this temperature.\cite{Morley2002}
There is no feature consistent with a rotational anomaly.  For
single crystal \#2, the behavior is significantly different than that
for the polycrystal, but reproducible.  In Fig.  \ref{Qinv}, four $T$
sweeps at 3 $\mu$m/sec are shown for single crystal \#2, two on
warming and two on cooling.  The damping is the same as that of the
polycrystal below 150 mK, but exceeds that of the
polycrystal in part of the upper $T$ range. There is a local maximum
in the damping near 250 mK, coincident with the maximum rate of
increase of $\delta f$, though the $T$ dependence is qualitatively
different for warming and cooling.  The warming and cooling curves
converge to the same damping level at 600 mK, closing the hysteresis
loop.  For single crystal \#1, $Q^{-1}(T)$ reaches a small
reproducible maximum at 200 mK on warming, which again coincides with
the maximum rate of increase in the corresponding measurement of
$\delta f$.  The four closed circles are the best fit values of the
dissipation for Lorentzians fitted to the resonant response of the TO,
as discussed in reference to Figs. \ref{raw} and \ref{df_rel}, validating the
accuracy of our measurements at a constant drive frequency.

\begin{figure}[ptb]
\centering
\graphicspath{{./figures/}}
\includegraphics[width=3.4in]{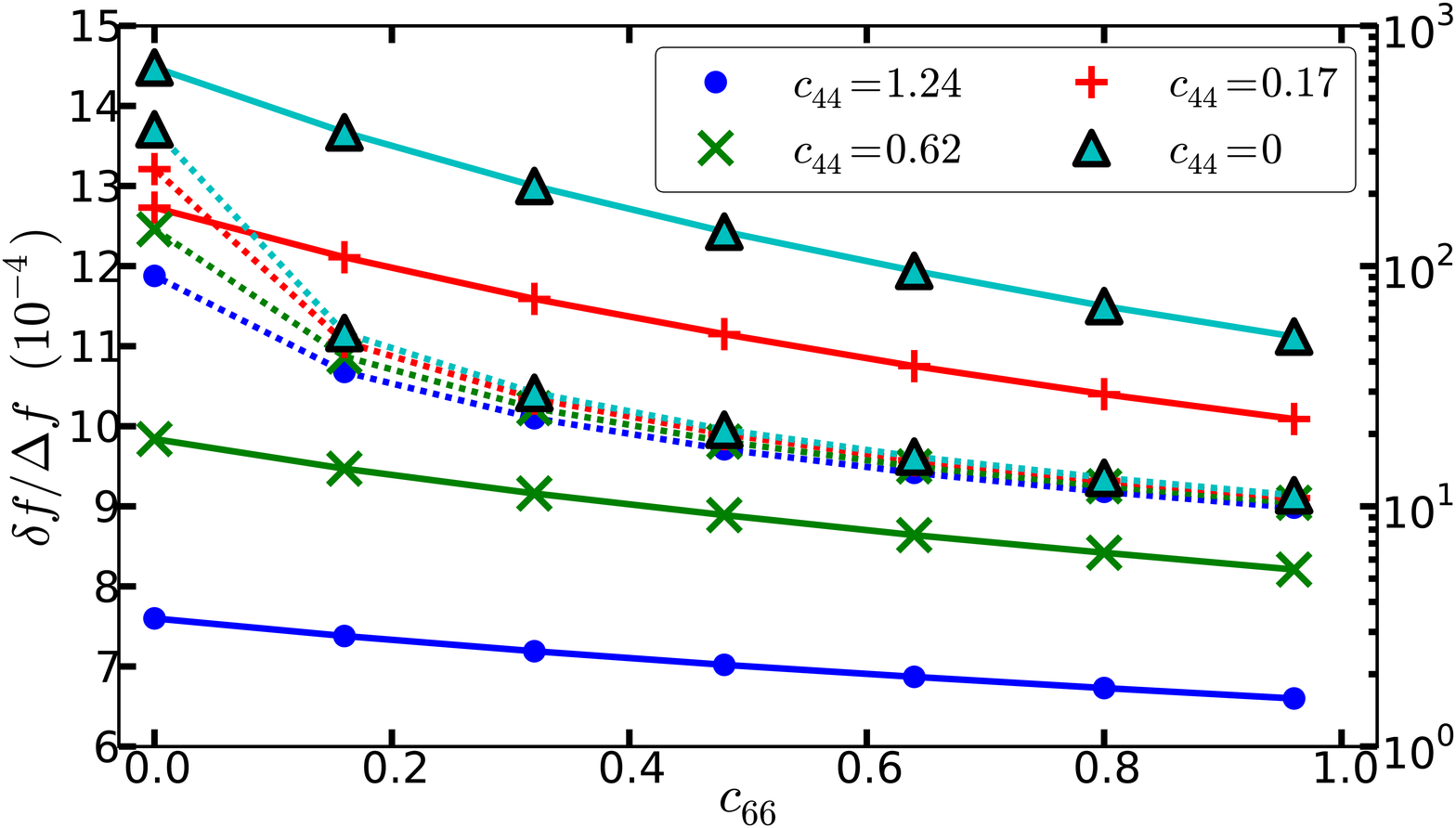}
\caption{(color) Change in the TO resonant frequency $\delta f$ relative to the mass loading $\Delta f$ as a function of the helium elastic constants $c_{44}$ and $c_{66}=(c_{11}-c_{12})/2$.\cite{Marisprivate}  The calculation is of the same type as in Ref. \onlinecite{Maris2011}, with a mesh spacing of 0.5 mm.  Solid (dashed) lines correspond to the $c$-axis being perpendicular (parallel) to the rotation axis of the TO and the left (right) vertical axis of the plot.  The elastic constants measured at high temperatures at which phonon damping prevents dislocation motion are $c_{44}=1.24$ and $c_{66}=0.96$.\cite{Greywall1977}  We assume that $\Delta
c_{12}=-\Delta c_{11}$, which is one way of satisfying the constraint that the strain energy decreases as the crystal softens.}
\label{assf}
\end{figure}

\section{Discussion}
Our observation of larger frequency shifts in single crystals than in
polycrystals does not seem consistent with supersolidity since the density
of dislocations along which the supersolid may flow is presumably at least
as large in polycrystals as in single crystals. Thus softening of the
crystals seems like the most likely explanation for our observations.\cite{Day2007,Zhou2011} One
usually assumes that edge dislocations easily glide only parallel to the
basal planes of the hcp structure \cite{Tsuruoka1979} and therefore that $%
c_{44}$ is the only elastic coefficient that softens in helium crystals. \
However, the magnitude of the frequency shift we observe in single crystal
\#2 is too large to be explained by variation of $c_{44}$ alone. According
to Ref.~\onlinecite{Maris2011}, the 86\% reduction in $c_{44}$ (relative to
its intrinsic, low $T$ value) observed in Ref. \onlinecite{Rojas2010} can produce a
frequency shift as large as 0.03\% of the mass loading. Single crystal \#2
had a three times larger shift of 2~mHz (0.1\% of the mass loading). If
dislocations can glide parallel to non-basal high density planes, other
elastic coefficients could soften, for example $c_{66}=(c_{11}-c_{12})/2$. \
Calculations of the TO frequency shift as a function of $c_{44}$ and $c_{66}$
are shown in Fig. \ref{assf}.\cite{Marisprivate} If the $c$-axis of the crystal is
parallel to the axis of rotation of our TO and $c_{44}$ is close to zero, $%
c_{66}$ would need to soften by approximately 50\% to produce a frequency
shift of 0.1\% of mass loading, as with single crystal \#2. It is also interesting to note that the
frequency shift reaches very large values as $c_{44}$ and $c_{66}$ approach
zero, for this orientation.

There are some interesting differences between the present results and
previous observations. Clark \textit{et al.} \cite{Clark2007} observed a
frequency shift that was three to five times larger when samples were grown
at constant volume (forming polycrystals) than at constant $T$ and $P$ from
the superfluid (presumably forming single crystals). This is the opposite
of the dependence on crystal quality that we observe, even though the
topology the cells in the two experiments is the same and the rigidity of
the cells is similar. This difference could be due to the different shapes
 of the two cells (the interior of our cell has no sharp angles so
as to minimize any liquid regions after growth at constant temperature) or
the fabrication material (our cell is the only one to be constructed with
polished sapphire, producing very smooth walls). We also observe the softening
in single crystals at a higher temperature than in other experiments,
except perhaps that of Penzev \emph{et al}.~\cite{Penzev2007} We checked that this
difference cannot be attributed to a difference in $^{3}$He content. A possible
explanation is that 
the geometry of our cell and our growth process produces crystals with very
low dislocation density.  As the dislocation density decreases, the average pinning length
in the absence of impurities, i.e. the network length $L_N$, increases.  This in turn causes
an increase in the pinning temperature $T_p\approx E_B/k_B \ln (a/xL_N)$ (supplement to Ref. \onlinecite{Day2007}),
where $E_B$ is the $^3$He impurity binding energy, $a$ is the atomic spacing along the dislocation, and $x$ is the bulk concentration
of $^3$He impurities. Thus the onset of the frequency shift should occur at a higher temperature.

Our results therefore suggest that our geometry, cell materials and crystal
growth procedure produce very high quality crystals. It would thus be
interesting to directly measure the dislocation density in the crystal. \
Our results also suggest that elastic coefficients other than $c_{44}$ may
change. In view of this, it would also be interesting to measure the
temperature dependence of all elastic coefficients of $^{4}$He oriented
single crystals, not only the average shear modulus of polycrystals~\cite%
{Day2007} or some combination of coefficients in single crystals.~\cite%
{Rojas2010,Mukharsky2009}

\section{Acknowledgements}
We gratefully acknowledge very helpful conversations with Humphrey Maris and John Beamish. This work is supported by the ERC grant
AdG247258-SUPERSOLID and by the NSF grants DMR~0706339 and DMR~1103159.

\bibliographystyle{apsrev}
\newif\ifabfull\abfulltrue

\end{document}